\documentstyle[twocolumn,prb,aps]{revtex}
\begin{document}
\draft
\tightenlines
\input epsf

\twocolumn [ \hsize\textwidth\columnwidth\hsize\csname @twocolumnfalse\endcsname
 
\title{Phase diagram of the extended Hubbard chain
with charge-dipole interactions.}
\author{M.E. Torio$^{a}$, A.A. Aligia$^{b}$, K. Hallberg$^{b}$ and H.A. Ceccatto$%
^{a} $.}
\address{
$^{a}$Instituto de F\'{i}sica Rosario, CONICET-UNR,\\
Bv 27 de Febrero 210 bis, (2000) Rosario, Argentina \\
$^{b}$Centro At\'{o}mico Bariloche and Instituto Balseiro,\\
Comisi\'on Nacional de Energ\'{\i}a At\'{o}mica, \\               
8400 Bariloche, Argentina.}

\maketitle

\thispagestyle{empty}
 
\vskip .5cm
 
\begin{abstract}
We consider a modified extended Hubbard model (EHM) which, in addition to
the on-site repulsion $U$ and nearest-neighbor repulsion $V$, includes
polarization effects in second-order perturbation theory. The model is
equivalent to an EHM with renormalized $U$ plus a next-nearest-neighbor
repulsion term. Using a method based on topological quantum numbers (charge
and spin Berry phases), we generalize to finite hopping $t$ the quantum
phase diagram in one dimension constructed by van den Brink {\it et al. }%
(Phys. Rev. Lett. {\bf 75}, 4658 (1995)). At hopping $t=0$ there are two
charge density-wave phases, one spin density-wave phase and one intermediate
phase with charge and spin ordering, depending on the parameter values. At $%
t\neq 0$ the nature of each phase is confirmed by studying correlation
functions. However, in addition to the strong-coupling phases, a small
region with bond ordering appears. The region occupied by the intermediate
phase first increases and then decreases with increasing $t$, until it
finally disappears for $t$ of the order but larger than $U$. For small $t$,
the topological transitions agree with the results of second order
perturbation theory.
 
\end{abstract}  


\vskip 1.cm
]

\section{Introduction}

Atomic polarizabilities are very important in solid state physics, since
they contribute to the screening of the effective on-site repulsion. For
example, from atomic data the bare on-site Coulomb energy for Cu$^{+2}$ is $%
U_{0}=E($Cu$^{+3})+E($Cu$^{+})-2E($Cu$^{+2})\sim 17$ eV, while constrained
density- functional calculations for La$_{2}$CuO$_{4}$ give $U\sim 10$ eV 
\cite{ann,hyb,gra}. In the case of BaBiO$_{3}$, while $U_{0}($Bi$^{+4})$ $%
\sim 11$ eV, it has been suggested that the effective $U$ becomes negative 
\cite{var}. Interesting effects of the polarizability also appear in the
physics of small clusters. In particular, while in neutral clusters of a few
atoms of Hg the only relevant interatomic interactions are van der Waals
forces (originated by fluctuations of the polarization), a transition to
covalent and finally to metallic bonding takes place as the size of the
cluster increases \cite{rad,bre,gar}. A very good fit of the observed
ionization potential as a function of cluster size \cite{rad} has been
obtained using a theory which includes charge-dipole and dipole-dipole
interactions in a two band model (representing the 6s and 5p electrons of
Hg). \cite{epl}

Recently, a simple one-band model including screening effects has been
proposed and studied.\cite{bri,mei} The main simplifying assumption is that
the electronic transitions responsible for the screening do not involve the
valence electrons (only transitions from one deep level to another one well
above the Fermi energy are considered). Furthermore, only nearest-neighbor
charge-dipole interactions are included. These facts renders the model more
amenable to many-body treatments, and interesting conclusions have been
extracted. In particular, in one dimension (1D), in addition to the usual
charge density wave (CDW1) and spin density wave (SDW) phases of the
extended Hubbard chain, two new phases appear in the ground state for
hopping $t=0$ \cite{bri,mei}: a charge density wave with a unit cell of four
lattice parameters (CDW2), and a combined spin and charge density wave with
a unit cell of six lattice parameters called intermediate (I) phase. The 1D
phase diagram for $t=0$ was constructed, and the energy gap, static
correlation functions and one-particle excitation spectra have been studied
numerically for a few cases with $t\neq 0.$ \cite{bri,mei} However, the
phase diagram for $t\neq 0$ has not been investigated yet.

The main difficulty in obtaining the phase diagram for $t\neq 0$ is that,
for finite systems, the conventional order parameters are smooth functions
of the parameters of the model, and the values at which the transitions take
place in the thermodynamic limit cannot be determined precisely. However,
recently it has been shown that in certain cases, including the usual
extended Hubbard model with repulsive interactions \cite{gag}, the
thermodynamic phases are characterized by different topological numbers
(determined by charge and spin Berry phases), and these jump suddenly for
any system size. \cite{gag,topo,epl2} Extrapolating the position of this
jump to the thermodynamic limit, a very accurate determination of phase
boundaries was obtained, as established by comparison with available exact
results in the strong-coupling limit and field-theory ones in the
weak-coupling case. \cite{topo,bos} In 1D, the method of the Berry phases is
related to the method of  crossings of excitacion levels based on the
conformal field theory with renormalization group analysis. \cite{japo,Naka}
This method (abbreviated as level-crossing method in the rest of the paper)
is not expected to work in the strong coupling limit.

In this work we use the method of Berry phases to calculate the phase
diagram of the model for $t\neq 0$. In particular, we show that in this case
all the thermodynamic phases can be characterized by the topological
numbers, including also a new phase with bond ordering previously suggested
in the literature.\cite{Naka} The nature of each phase is confirmed by
calculating different correlation functions. In Section II we describe the
model and the strong-coupling thermodynamic phases. Section III contains a
brief description of the calculation of the topological numbers and the
values they take for the $t=0$ phases. In Section IV we present the quantum
phase diagram for several values of $U_{0}/t$, compare it with the results
of perturbation theory in $t$, and discuss the behavior of the correlation
functions. In Section V we summarize our results and discuss the extent and
limitations of the method of the topological transitions and that of the
level crossings.

\section{Model and $t=0$ ground-state phases}

The model is defined by the Hamiltonian

\begin{equation}
H=H_{0}+H_{cd},  \label{h1}
\end{equation}
where $H_{0}$ is the usual extended Hubbard chain: 

\begin{eqnarray}
H &=&-t\sum_{i\sigma }(c_{i+1\sigma }^{\dagger }c_{i\sigma }+{\rm H.c.}%
)+U_{0}\sum_{i}n_{i\uparrow }n_{i\downarrow }\;
\nonumber \\
&&+\;V\sum_{i}n_{i}n_{i+1},
\label{h0}
\end{eqnarray}

and $H_{cd}$ describes charge-dipole interactions (dipole-dipole
interactions are neglected):

\begin{equation}
H_{cd}=-\sum_{i}{\bf p}_{i}\cdot {\bf E}_{i}.  \label{hcd}
\end{equation}
Here ${\bf p}_{i}$ and ${\bf E}_{i}$ are the dipole moment operator and
electric field at site $i$ respectively. Treating this term in second-order
perturbation theory\cite{gar,epl,bri,mei} it takes the form $-\frac{1}{2}%
\sum_{i}\alpha _{i}E_{i}^{2}$, where $\alpha _{i}$ is the polarizability at
site $i$. Considering only the contribution to the electric field coming
from nearest-neighbor atoms, and assuming that $\alpha _{i}$ is independent
of the site and its occupancy, one can write:

\begin{equation}
H_{cd}=-P\sum_{i}(n_{i+1}-n_{i-1})^{2},  \label{hcd2}
\end{equation}
where $P=\alpha e^{2}/(2a^{4})$ and $a$ is the lattice parameter. Expanding $%
H_{cd}$, the total Hamiltonian can be cast into the form:

\begin{eqnarray}
H &=&-t\sum_{i\sigma }(c_{i+1\sigma }^{\dagger }c_{i\sigma }+{\rm H.c.}%
)+(U_{0}-4P)\sum_{i}n_{i\uparrow }n_{i\downarrow }\;
\nonumber \\
&&+\;V\sum_{i}n_{i}n_{i+1}
+2P\sum_{i}n_{i}n_{i+2}-2P\sum_{i}n_{i}.  \label{hl}
\end{eqnarray}

The last term merely renormalizes the chemical potential. As stated before 
\cite{bri,mei}, within the above mentioned approximations $H_{cd}$ reduces
the on-site interaction and generates a next-nearest-neighbor repulsion.

We are interested in the half-filled case of one particle per site. For $t=0$
the model can be solved exactly \cite{bri,mei}, and at this filling four
different phases appear in the ground state depending on the parameter
values. We use the notation $\bigcirc $, $\bullet $,$\uparrow $, and $%
\downarrow $, to represent empty, doubly occupied, and singly occupied sites
with spin up and down respectively. If $V$ is much larger than the other
interactions, nearest-neighbor occupancy must be avoided and the ground
state can be represented as ($...\bigcirc \bullet \bigcirc \bullet \bigcirc
\bullet \bigcirc \bullet ...$). This state is denoted as charge density wave
1 (CDW1). If $U_{0}$ is the dominant interaction all sites are singly
occupied, and an infinitesimally small $t$ is enough to introduce
antiferromagnetic correlations between nearest-neighbors. Thus, we represent
this spin density wave (SDW) as ($...\uparrow \downarrow \uparrow \downarrow
\uparrow \downarrow ...$), although there is no true spin long-range order
in one dimension. If $P$ is very large, occupancy at next nearest neighbors
is avoided and the ground state (CDW2) is ($...\bigcirc \bigcirc $ $\bullet
\bullet \bigcirc \bigcirc $ $\bullet \bullet ...$). Finally, for
intermediate values of the parameters (for example $V=0.5U_{0}$, $%
P=0.15U_{0} $) there is an intermediate (I) phase, the unit cell of which
can be represented as $\bullet \uparrow $ $\bigcirc \bullet \downarrow
\bigcirc $.

\section{Berry phases}

We have calculated the ground state $|g(\Phi _{\uparrow },\ \Phi
_{\downarrow })\rangle $ of the model in rings of different lengths $L$,
threaded by fluxes $\Phi _{\sigma }$ for spin $\sigma $ (the hopping for
each spin is affected by a factor $e^{i\Phi _{\sigma }/L}$ as described in
detail elsewhere \cite{gag,topo,epl2}). We call charge (spin) Berry phase $%
\gamma _{c}$ ($\gamma _{s}$), the phase captured by the ground state when it
is followed adiabatically in the cycle $0\leq \Phi \leq 2\pi $, keeping $%
\Phi _{\uparrow }=\Phi _{\downarrow }=\Phi $ ($\Phi _{\uparrow }=-\Phi
_{\downarrow }=\Phi $). Discretizing the interval $0<\Phi <2\pi $ into $N$
points $\Phi _{i}=2\pi i/N$, the Berry phases are calculated numerically by
the gauge-invariant expression \cite{gag}: 

\begin{eqnarray}
\gamma _{c,s} &=&-\lim_{N\rightarrow \infty }\{\text{Im}[\ln (\Pi
_{i=0}^{N-2}\langle g(\Phi _{i},\pm \Phi _{i})|g(\Phi _{i+1},\pm \Phi
_{i+1})\rangle  \nonumber \\
&&\langle g(\Phi _{N-1},\pm \Phi _{N-1})|g(2\pi )\rangle )]\},  \label{be}
\end{eqnarray}  

\noindent where $|g(2\pi )\rangle $ represents $|g(2\pi ,\pm \ 2\pi )\rangle 
$ obtained directly from $|g(0,\ 0)\rangle $: 

\begin{equation}
|g(2\pi )\rangle =\exp [i{\frac{{2\pi }}{L}}\sum_{j}j{(}n_{j\uparrow }\pm
n_{j\downarrow })]|g(0,0)\rangle .  \label{be2}
\end{equation}

An important property of the charge Berry phase is that if the system is
modified by some perturbation, the change in the polarization $P_{\uparrow
}+P_{\downarrow }$ is proportional to the corresponding change in $\gamma
_{c}.$\cite{ort} Here $P_{\sigma }$ is the contribution of electrons with
spin $\sigma $ to the polarization of the system. Similarly, changes in $%
\gamma _{s}$ are related to changes in the difference $P_{\uparrow
}-P_{\downarrow }$ between the electric polarizabilities for spin up and
down \cite{epl2}:

\begin{equation}
\Delta P_{\uparrow }\pm \Delta P_{\downarrow }=e\Delta \gamma _{c,s}/2\pi 
\text{ }[{\rm mod}(e)].  \label{po}
\end{equation}
A crucial property for our purposes is that, in systems with inversion
symmetry like ours, $\gamma _{c}$ and $\gamma _{s}$ can only be either $0$
or $\pi $ [{\rm mod}$(2\pi )$] (the argument of the logarithm in Eq. (\ref
{be}) turns into its complex conjugate under inversion). Then, the Berry
phase vector ${\bf \gamma }=(\gamma _{c},\gamma _{s})$ cannot vary
continuously, and a jump in ${\bf \gamma }$ corresponds to a transition in
at least one of the topological numbers $\gamma _{c}/\pi $, $\gamma _{s}/\pi 
$. Our main hypothesis is that the topological transitions correspond to
phase transitions in the thermodynamic limit. This has been confirmed by
different numerical and analytical studies in the Hubbard model with
correlated hopping. \cite{topo,bos} In the present case, we check this by
comparing different correlation functions (see next section). In addition,
it has been shown that the opening of a spin gap in a Luttinger liquid phase
of a spin rotational invariant model is accompanied by a topological
transition in $\gamma _{s}$ \cite{epl2}.

Let us discuss the values of ${\bf \gamma }$ for the four possible ground
states of Eq.(\ref{h1}) in the half-filled system with $t=0$, discussed at
the end of the previous section. For $t=0$, we can choose a gauge in which
all scalar products in Eq. (\ref{be}) are equal to 1, except eventually the
last one. Then the Berry phases are defined by the argument of the
exponential in Eq. (\ref{be2}). If the ground state is the state ($%
...\bigcirc \bullet \bigcirc \bullet \bigcirc \bullet \bigcirc \bullet ...$)
of the CDW1, then this argument is clearly $0$ for $\gamma _{s}$, while for $%
\gamma _{c}$ it is equal to $i{\frac{{2\pi }}{L}}\sum_{j}j{(}n_{j\uparrow
}+n_{j\downarrow })$ $=i{\frac{{2\pi }}{L}}\sum_{m=1}^{L/2}4m$ $=i2\pi
(L+2)\equiv 0$ [{\rm mod}$(2\pi i)$]. The same result is obtained for the
other CDW1\ state [$...\bullet \bigcirc \bullet \bigcirc \bullet \bigcirc
\bullet \bigcirc ...$], so ${\bf \gamma }$(CDW1)$=(0,0).$ One can construct
a N\'{e}el state (representing the SDW) displacing half of the electrons
(for example those with spin up) one lattice parameter either to the left or
to the right. Then, from Eqs. (\ref{po}) one has ${\bf \gamma }$(SDW)$=(\pi
,\pi )$. This result can also be easily obtained calculating the exponent of
(\ref{be2}), and is not altered by spin fluctuations. \cite{gag,topo} For
the CDW2 with order ($...\bigcirc \bigcirc $ $\bullet \bullet \bigcirc
\bigcirc $ $\bullet \bullet ...$) clearly $\gamma _{s}=0$, because the spin
at each site is compensated. For $\gamma _{c}$ one has ${\frac{{2\pi }}{L}}%
\sum_{j}j{(}n_{j\uparrow }+n_{j\downarrow })$ $={\frac{{2\pi }}{L}}%
\sum_{m=1}^{L/4}(16m-2)$ $=2\pi (L+4-1/2)\equiv \pi $ [{\rm mod}$(2\pi )$].
Then, ${\bf \gamma }$(CDW2)$=(\pi ,0).$ While the phases above have been
calculated for $t=0$, it is clear that by continuity each thermodynamic
phase retain the same value of ${\bf \gamma }$ unless a topological
transition takes place.

The vector ${\bf \gamma }$ in the I phase cannot be calculated as easily as
above for $t=0$. The reason is that the state ($...\bullet \uparrow $ $%
\bigcirc \bullet \downarrow \bigcirc ...$) does not have a definite parity
under inversion and other states related with it by inversion have different
exponents of Eq. (\ref{be2}). For $t\neq 0$ the ground state should have a
well defined parity, and ${\bf \gamma }$ depends on the particular linear
combination of the above mentioned states which constitutes the ground
state. Numerically we found ${\bf \gamma }$(I)$=(0,\pi )$. This is a
fortunate fact, since then at $t=0$ the four thermodynamic phases are
characterized by all the possible values of ${\bf \gamma }$. Thus, the
corresponding phase boundaries can be determined from topological
transitions. Notice, however, that this does not rule out the possibility
that at $t\neq 0$ new phases having the same ${\bf \gamma }$ values could
appear. We will show in the next section that this is in fact what happens.
To summarize and for future use, the Berry phase vectors for each of the
strong-coupling phases are ${\bf \gamma }$(CDW1)$=(0,0)$, ${\bf \gamma }$%
(SDW)$=(\pi ,\pi )$, ${\bf \gamma }$(CDW2)$=(\pi ,0)$, and ${\bf \gamma }$(I)%
$=(0,\pi )$.

\section{Numerical results}

In this section we present the ground-state phase diagram of model (\ref{h1}%
) for $t\neq 0.$ As discussed in the previous section, it has been obtained
by studying the behavior of the Berry phases in rings of $L=8$ and $12$
sites. Notice however that the order corresponding to the I phase cannot be
accommodated in rings of $8$ sites because of its $6$-site unit cell. In
Fig. 1 we show the phase diagram for $U_{0}/t=10$, $3$ and $1$, which
correspond to the strong to moderate coupling regimes. As can be seen in
Fig. 1a, for $U_{0}/t=10$ the results are similar to the strong limit case ($%
t=0$). Moreover, there are no sensible differences between the values
obtained in rings of 8 and 12 sites, which indicates that they are
representative of the thermodynamic limit results. Fig. 1b shows that
already for $U_{0}/t=3$ the behavior departs from the one corresponding to
the strong coupling limit; in particular, there are stronger finite-size
effects in the transition lines CDW1-I and CDW2-I, especially when the
transition line is determined through the jump in the spin Berry phase. The
line CDW2-I is better determined by the changes in the charge Berry phase,
which shows no major finite-size effects. We expect that, for this line,
going to larger systems the jumps in the spin Berry phase will coincide with
those of the charge Berry phase.

Note that with a small increase in $t$ from the strong coupling limit, the I
phase first increases with respect to the CDW1 and CDW2 phases. This can be
understood from perturbation theory in $t$. The energies of the different
phases calculated up to second order in $t$ are the following:

\[
\smallskip E(\text{SDW})=N\ \left( V+2P-\frac{4t^{2}\ln 2}{U_{0}-4P-V}%
\smallskip \right) , 
\]

\smallskip

\[
E(\text{CDW1})=N\ \left( \frac{U_{0}}{2}+2P-\frac{2t^{2}}{3V-U_{0}}%
\smallskip \right) , 
\]

\smallskip

\[
E(\text{CDW2})=N\ \left( \frac{U_{0}}{2}-2P+V-\frac{t^{2}}{12P-V-U_{0}}%
\smallskip \right) , 
\]

\smallskip

\begin{equation}
E(\text{I})=\frac{N}{3}\ \left[ U_{0}+2V-t^{2}\left( \frac{1}{P}+\frac{2}{%
8P+V-U_{0}}\right) \smallskip \right] .  \label{perturb}
\end{equation}
The phase diagram that results from these energies is represented in Fig. 2.
For $U_{0}/t=3$, the second order expression for $E($SDW$)$ is no more valid
(it gives a very low energy). Thus, we have not drawn the boundaries of the
SDW phase for this value of $U_{0}/t$. Instead, the other boundaries of the
I phase, and in general all boundaries for $U_{0}/t=10$ (except when some
denominator becomes of the order of $t$), agree very well with the
topological transitions represented in Fig. 1.

 
\vspace{-0.3cm}
\begin{figure}[ht]
\epsfxsize=8.cm
\epsfbox{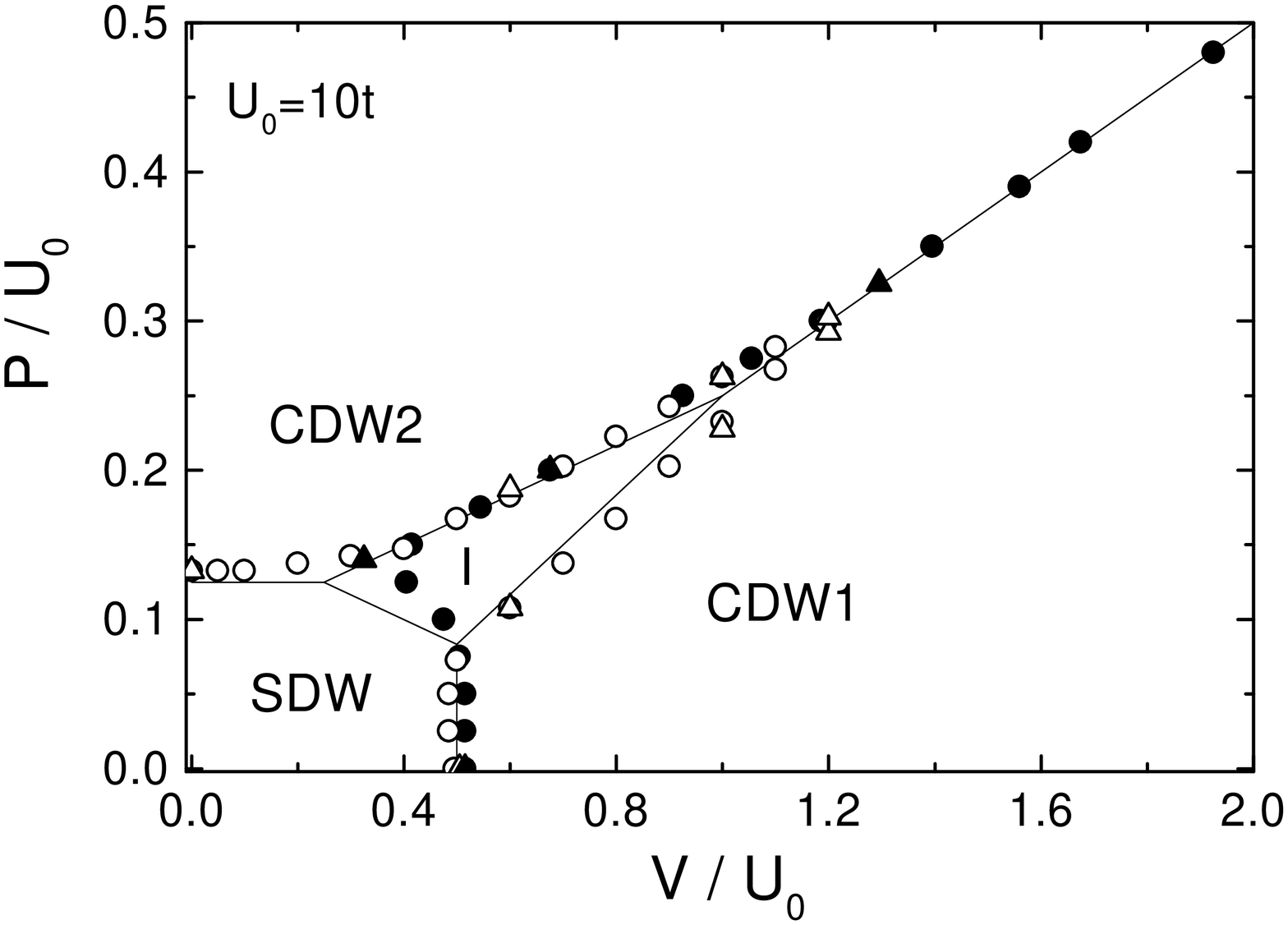}
\end{figure}
 
\vskip -0.95cm
\begin{figure}[ht]
\epsfxsize=8.cm
\epsfbox{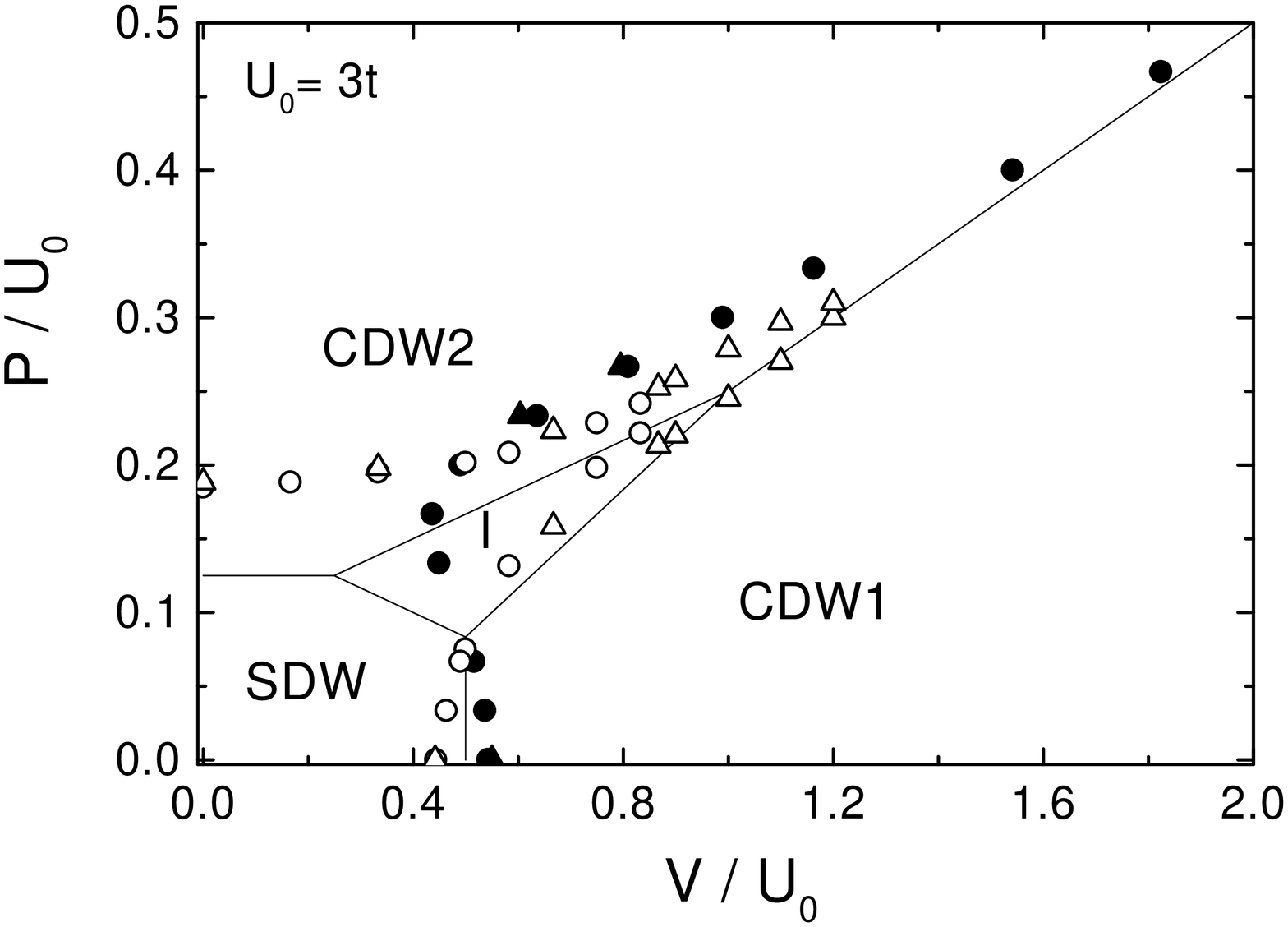}
\end{figure}
 
\vskip -0.95cm
\begin{figure}[ht]
\epsfxsize=8.cm
\epsfbox{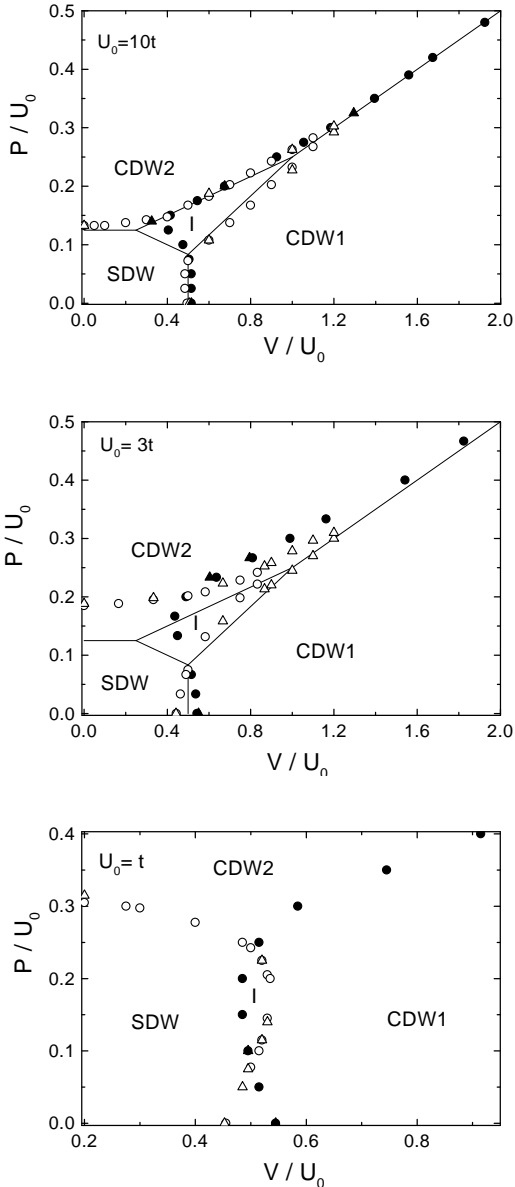}
\caption{Quantum phase diagram of model (\ref{h1}). a) $U_{0}/t=10$;
 b) $U_{0}/t=3$%
; c) $U_{0}/t=1$. Solid dots and triangles correspond to jumps in the charge
Berry phase obtained by considering rings with $L=8$ and $12$ sites
respectively; open dots and triangles indicate jumps in the spin Berry phase
for systems of the same length. The thin lines correspond to the $t=0$ phase
diagram.}
\end{figure}                   


From Eqs. (\ref{perturb}) and
Fig. 2, we see that near the points where the CDW1, CDW2 and I phases meet
for $t=0$ the stability of the I phase increases with $t$. This is due to
the fact that the electrons of the doubly occupied sites of the I phase can
jump to the neighboring singly occupied sites with small energy cost in $P$,
and without increasing on-site and nearest-neighbor repulsion energies. As a
consequence, for small $t\neq 0$, the electrons in the I phase become more
mobile and gain more kinetic energy than in the other two phases.

\smallskip On the other hand, for $U_{0}/t=3$ the topological transitions
seem to indicate that a new phase appears between the SDW and CDW1 phases
(see Fig. 1). This was only slightly visible for $U_{0}/t=10$ and is out of
the reach of the perturbative calculations. We wave studied this possibility
in more detail by considering the scaling of critical points on the axis $%
P=0 $ shown in Fig. 3. From this figure we see that the jumps in the spin
and charge Berry phases will not coincide even in the thermodynamic limit;
this fact is pointing to the emergency of a new order termed bond ordering
wave (BOW), as suggested in a recent study of the phase diagram of the EHM.%
\cite{Naka}


\vskip -0.4cm
 
\begin{figure}[ht]
\epsfxsize=8.5cm
\epsfbox{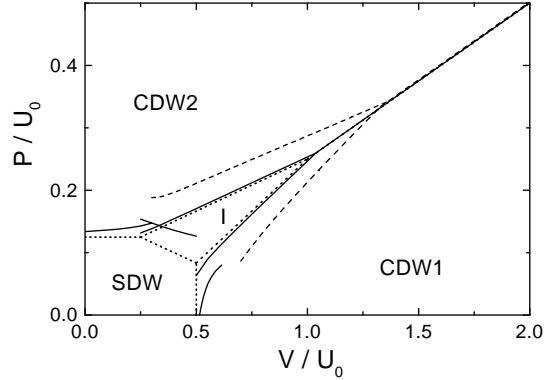}
\caption{ Phase boundaries obtained by perturbation theory in $t$. Full line $%
U_{0}/t=10$, dashed line $U_{0}/t=3$. The dotted line is the known result
for $t=0$. \cite{bri,mei}.}
\end{figure}
For $U_{0}/t=1$ (see Fig. 1c) the phase diagram looks quite different to the
strong coupling case, with the I phase shrinking from a triangular-like
shape to a smaller biconvex region. Furthermore, the BOW phase remains
present without much change. Notice that using bosonization (valid in the
extreme weak-coupling regime $t\rightarrow \infty $), the BOW phase is
absent for $P=0,$\cite{bos,jap} and the I and CDW2 phases should also
disappear since the expected instabilities have wave vector $2k_{F}=\pi $.

\vskip -0.4cm
\begin{figure}[ht]
\epsfxsize=8.5cm
\epsfbox{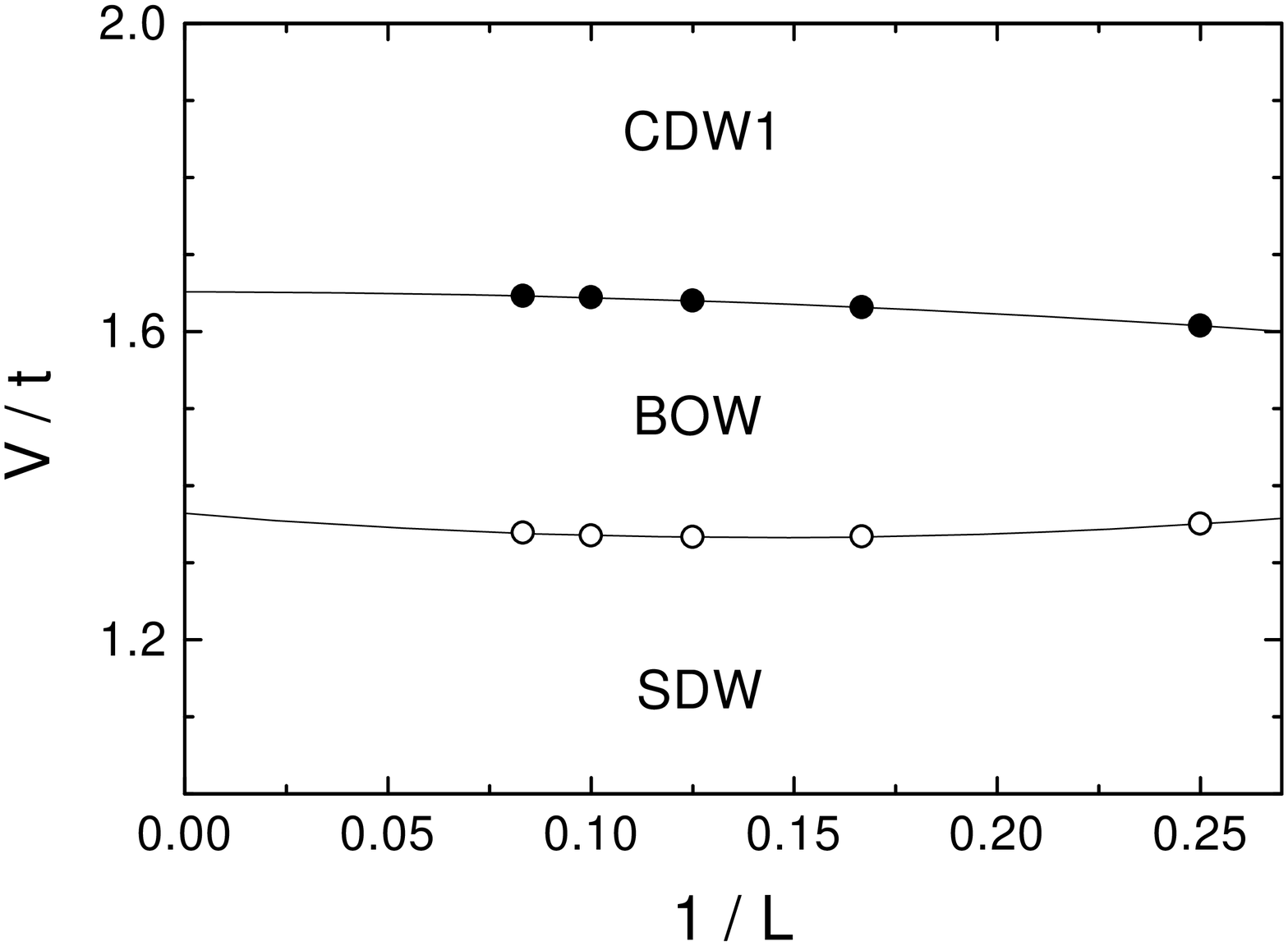}
\caption{Values of $V/t$ at which the charge (solid dots) and spin (open dots)
Berry phases jump on the $P=0$ axis, as a function of the ring length $L.$
We have taken $U_{0}/t=3.$}
\end{figure}                 

In order to have a better characterization of the order present in the
different phases, we have also computed static correlation functions along
the $P=0$ axis from $V/t=1$ to $V/t=2$, that is, traversing across the SDW,
BOW and CDW1 phases. Fig. 4 shows the spin and charge correlation functions
at the largest distance $L/2=5$ for a 10-site chain, and also the
correlation function corresponding to the expected order in the BOW phase, $%
C_{{\rm BOW}}(L/2)=\left\langle [O_{{\rm B}}(L/2)-\overline{O_{{\rm B}}]}[O_{%
{\rm B}}(0)-\overline{O_{{\rm B}}]}\right\rangle $, with $O_{{\rm B}%
}(j)=(-)^{j}\sum_{\sigma }(c_{\sigma ,j+1}^{\dagger }c_{\sigma ,j}+{\rm H.c.}%
).$ The vertical dotted lines correspond to the transition points
extrapolated in Fig. 2, and we can see the expected behavior for these three
functions along the chosen path.

\begin{figure}[ht]
\epsfxsize=13.5cm
\epsfbox{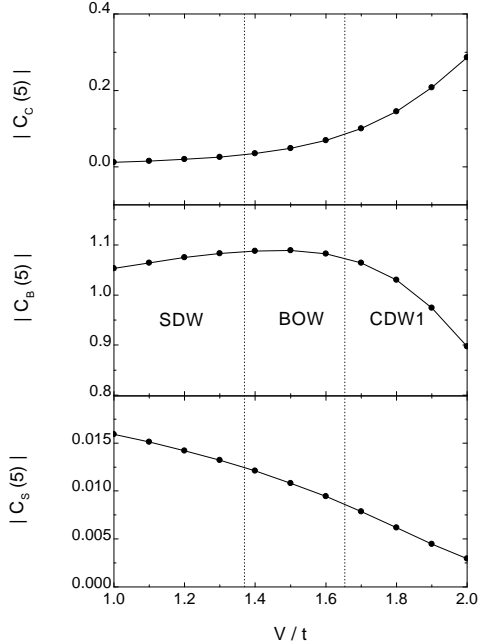}
\caption{Static charge ($C_{{\rm C}}$), BOW ($C_{{\rm B}}$) and
spin ($C_{{\rm S}}$
correlation functions at the largest distance $L/2=5$ for a 10-site ring.
The dotted vertical lines indicate the extrapolated points in Fig. 3.}
\end{figure}

Furthermore, in Fig. 5 we plot the
fluctuations of the three order parameters [for example $\chi _{{\rm BOW}}=%
\frac{1}{L^{2}}\sum_{ij}e^{iq(R_{i}-R_{j})}\langle O_{{\rm B}}(i)O_{{\rm B}%
}(j)\rangle $] at their peak values (corresponding to $k=\pi $)$,$ as a
function of the ring length and for $V/t=1.5,$ that is, inside the BOW
phase. Both the spin and charge fluctuations diminish with increasing
lengths, while the BOW fluctuations increase and most likely dominate in the
thermodynamic limit. The same analysis in the SDW and CDW1 phases is not so
conclusive because the BOW fluctuations do not clearly decay when increasing
the size of the system; however, since the orders present for small and
large values of $V/t$ are not a matter of discussion, this must be only due
to the small size of the systems considered.

 
\begin{figure}[ht]
\epsfxsize=9.5cm
\epsfbox{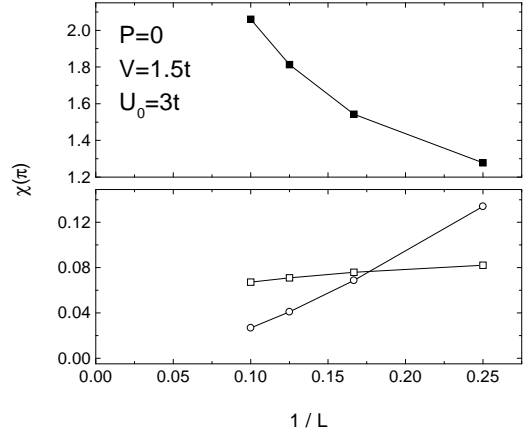}
\caption{Charge (open dots), spin (open squares) and BOW (solid dots)
order-parameter fluctuations at $k=\pi $ as a function of the
ring length $%
L. $ We have considered the parameters values $P=0$, $V/t=1.5$
and $U_{0}/t=3
$, which correspond to a point well inside the BOW phase.}
\end{figure}

\section{Discussion}

We have studied the effect of electronic polarization $P$ on the phase
diagram of the 1D extended Hubbard model. Following Refs.\cite{bri,mei}, we
have neglected the screening effect of valence electrons, although it was
important to explain experiments in Hg clusters \cite{gar,epl}. At $P=0$,
for a bipartite lattice in any dimension two phases are expected: a SDW for
large on-site Coulomb repulsion $U_{0}$ (without true long-range order in
1D), and a CDW1 for large nearest-neighbor repulsion $V$. In 1D the results
of Ref. \cite{Naka} and ours support the existence of a narrow region of
long-range bond ordering between the other two phases, if the hopping $t\neq
0$. Previous studies in 1D \cite{bri,mei} have shown that for small $t$ and
sufficiently large $P$, two new phases appear: CDW2 and I, with unit cells
containing four and six lattice parameters respectively, as described in
section II.

For $t=0$ it is clear (see section II) that a transition from one phase to
the other is accompanied by charge and /or spin transport. Since the charge
Berry phase $\gamma _{c}$ is a measure of the polarization of the system,%
\cite{ort} and similarly a jump in the spin Berry phase $\gamma _{s}$
indicates a jump in the difference in polarization for spin up and down,\cite
{epl2} it seems appropriate to determine the phase diagram from the jumps in
the topological numbers $\gamma _{c}/\pi $, $\gamma _{s}/\pi .$\cite{topo}
Using this method, we find that for fixed $U_{0}$, $V$ and $P$, the regions
occupied by the CDW2 and particularly the I phase, are reduced for
sufficiently large $t$. This result is expected on general grounds, since
only instabilities with wave vector $\pi $ are expected in the weak coupling
limit ($t\rightarrow \infty $), which should be well described by continuum
limit field theory.\cite{bos,jap,voit} For small $t$ (up to $U_{0}/3 $)
 the region occupied by the I phase first increases with $t$, and this
effect can be understood using second-order perturbation theory in $t$.

Although in the present case we succeded in determining all phase boundaries
from the jumps in the topological numbers $\gamma _{c}/\pi $, $\gamma
_{s}/\pi $, there are some cases in which the thermodynamic phases are
characterized by the same topological numbers, and the phase transition
cannot be detected. In other cases some topological transitions do not have
an obvious physical meaning.\cite{epl2,z} An example of the former is the
transition between SDW and triplet superconductivity{\bf \ }which takes
place for positive $U$ and negative $V$ in the extended Hubbard model in 1D,
which is however well determined by the level-crossing method.\cite{Naka}
This method, as used in Refs. \cite{japo,Naka}, is limited to 1D and is
based on the fact that for critical systems, due to conformal invariance,
the exponent of the distance dependence of different correlation functions
is given by the corresponding excitation energies.\cite{japo} For systems at
(or slightly out of) criticality, described by the sine-Gordon model at low
energies, renormalization arguments were used and logarithmic finite-size
corrections can be eliminated using appropriate combination of excitation
energies.\cite{japo,ian,karen} This method cannot be applied to phases which
are out of the reach of the continuum-limit field theory like the CDW2 and I
phases, or when there is phase separation. Instead, in the strong-coupling
limit ($t\rightarrow 0$) it is usually very simple to calculate the Berry
phases, even in some phase separated states \cite{epl2,z} (it reduces to a
simple counting argument, as shown in Section II), and the method of the
topological numbers is very precise. It is interesting to note that, in
spite of their different meaning and expected limitations, both methods give
the same results for transitions among the SDW, CDW1 and BOW phases. The
method of Berry phases has also been applied to 2D transitions \cite{ort}.

\section*{Acknowledgments}

One of us (AAA.) thanks J. Lorenzana for useful discussions. MET and HAC.
acknowledge support from CONICET. AAA. is partialy supported by CONICET.
This work was sponsored by PICTs 03-00121-02153 and 03-03833 of ANPCyT and
PIP 4952/96 of CONICET.

\end{document}